\documentclass[twocolumn,showkeys,preprintnumbers,amsmath,amssymb]{revtex4} 

\usepackage{graphicx}
\usepackage{dcolumn}
\usepackage{bm}

\begin{document}

\title{Geometrical Nonlinearity of Circular Plates and Membranes: \\
       an Alternative Method}
\author{ D. Cattiaux$^{*}$, S. Kumar$^{*}$, X. Zhou$^{**}$, A. Fefferman$^{*}$ and E. Collin$^{*,\dag}$}

\address{(*) Univ. Grenoble Alpes, Institut N\'eel - CNRS UPR2940, 
25 rue des Martyrs, BP 166, 38042 Grenoble Cedex 9, France \\
          (**) IEMN, Univ. Lille - CNRS UMR8520, 
Av. Henri Poincar\'e, Villeneuve d'Ascq 59650, France 
}

\date{\today}

\begin{abstract}

We apply the well-established theoretical method developed for geometrical nonlinearities of micro/nano-mechanical clamped beams to circular drums. 
The calculation is performed under {\it the same} hypotheses, the extra difficulty being to analytically describe the (coordinate-dependent) additional stress generated in the structure by the motion. 
Specifically, the model applies to {\it non-axisymmetric} mode shapes.
An analytic expression is produced for the Duffing (hardening) nonlinear coefficient, which requires {\it only} the knowledge of the mode shape functions to be evaluated.
This formulation is simple to handle, 
 and does not rely 
on complex numerical methods. 
Moreover, no hypotheses are made on the drive scheme and the nature of the in-plane stress: it is not required to be of electrostatic origin. 
%
We confront our predictions with both typical experimental devices and relevant theoretical results from the literature.
%
Generalization of the presented method to Duffing-type mode-coupling should be a straightforward extension of this work. 
We believe that the presented modeling will contribute to the development of nonlinear physics implemented in 2D micro/nano-mechanical structures.
\end{abstract}


\maketitle

\section{Introduction}

The field of micro- and nano- electro-mechanics (MEMS and NEMS) \cite{roukescleland,clelandbook,schmidbook} has been continuously expanding over the last decades. 
These devices, which transduce motion into electrical signals, have been both developed into sensors (e.g. pressure gauge \cite{ekinci}) and components (e.g. r.f. signal mixer \cite{purcell}).
Beyond the notorious accelerometer \cite{acceleroNature} and mass spectroscopy \cite{roukesmass} applications, it even becomes possible today to embed nanomechanical elements into {\it quantum} electronic circuits \cite{cleland2010,quantelecsimmonds,quantelec2}.

Within the field, nonlinearities can be both a limitation or a {\it resource}. 
For all systems that build on linear response, nonlinearities of all kinds limit the dynamic range of the device \cite{roukesdynrange}.
On the other hand, one can devise efficient schemes that rely on nonlinearities to work: this very rich area includes applications such that e.g. amplification of small signals \cite{buksgain}, bit storage \cite{yamaguchi,warner}, and synchronization of oscillators \cite{crosssync,Matheny} among others.

In both cases, understanding and {\it mastering} the sources of nonlinearities is required, in order to tailor them on demand: maximizing, or minimizing them \cite{kozinsky,kacem,turner,defoort}. 
The main feature that impacts the dynamics of MEMS/NEMS is a {\it Duffing-type} nonlinear behavior \cite{bush,crosslifshitzbook}. The basic modeling capturing the physics is a $\tilde{k}\, x^3$ restoring force inserted in the dynamics equation of the mechanical mode; in practice, other terms may also contribute and be taken into account \cite{crosslifshitzbook,PRBEddy}.

Even if the materials are perfectly Hookean, all devices experience nonlinear behavior at large deformations: these arise from purely {\it geometrical} considerations. For flexural doubly-clamped beams, it consists of the extra stress stored in the beam under motion because of stretching \cite{crosslifshitzbook}. This effect has been widely studied experimentally, even beyond the nonlinear features of a single mode: the same effect indeed couples all the flexural modes of the structure \cite{kunalNJP,NanoRoukes,venstra,Olive}. 
The measurements are in very good agreement with the simple stretching theory, that can be found e.g. in Ref. \cite{crosslifshitzbook}. \\

		\begin{figure}
		\centering
		\vspace*{-1.cm}
	\includegraphics[width=9cm]{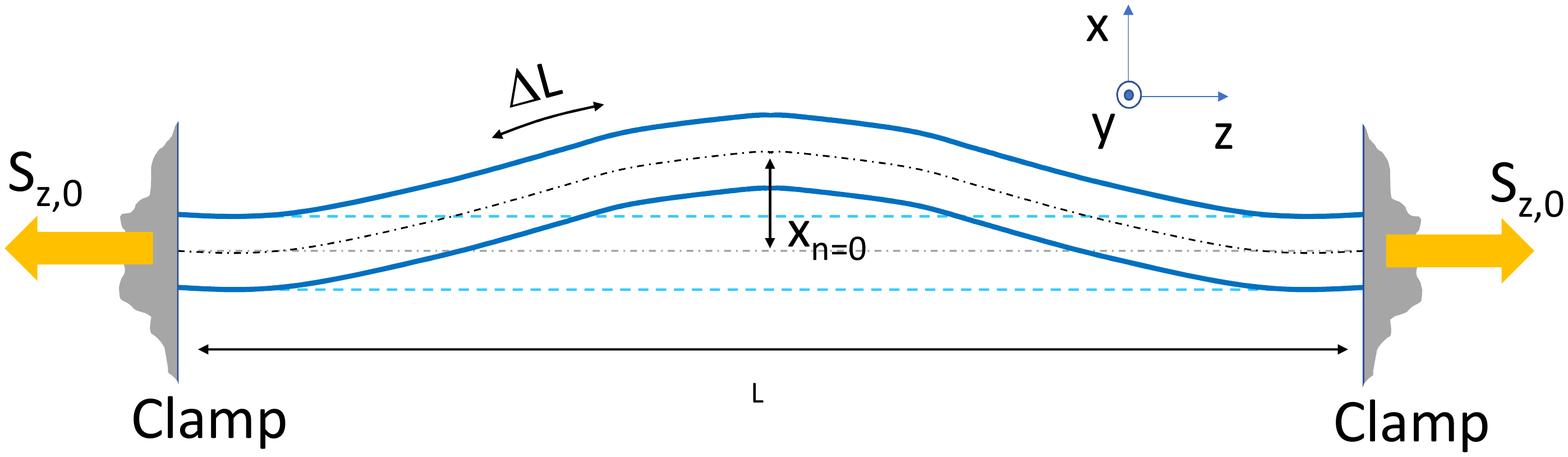}
	\vspace*{-2cm}
			\caption{
			(Color online) Schematic of a doubly-clamped device, in its fundamental flexure ($n=0$ mode). The axial force load $S_{z,0}$ is here tensile.}
			\label{fig_scheme}
		\end{figure}
			
For these reasons, we propose to extend this modeling to the 2D case of a drum. 
Our aim is to produce an {\it analytic, robust and simple} expression for the geometric nonlinear Duffing coefficient, similar to the 1D solution.
The modeling remains at a {\it generic level}, not introducing any specific drive fields, and applies to {\it non-axisymmetric} modes as well as to axisymmetric ones.
We shall start in Section \ref{beambase} by reviewing the beam nonlinear mathematics, and discuss the concepts and limitations of this approach.
In Section \ref{problem}, we present the adaptation of it to a 2D axisymmetric geometry; the stress field is discussed in Section \ref{stressfield} and the solution is given in Section \ref{soluce}.
Our results are discussed in Section \ref{discuss}, with a comparison to both experiments and theory from the literature.

%
\section{Beam theory basis}
\label{beambase}

Let us start by recalling the basics of the geometrical nonlinear modeling of clamped beams.
We first write the Euler-Bernoulli equation that applies to thin-and-long structures \cite{bernouille,timo}:
\begin{equation}
E_z I_z \frac{\partial^4 f( z,t ) }{\partial z^4} + S_z \frac{\partial^2 f( z,t ) }{\partial z^2} = - \rho A_z \frac{\partial^2 f( z,t ) }{\partial t^2}, \label{EB}
\end{equation}
with $E_z$ the Young's modulus, $I_z$ the second moment of area, $S_z$ the axial force load, $\rho$ the mass density and $A_z$ the section area. The $z$ index refers to the axis pointing along the beam, see Fig. \ref{fig_scheme}. The beam is assumed homogeneous with a constant cross section over its length $L$.
The function $f( z,t )$ describes the transverse motion of the structure (in the $x$ direction), with the proper boundary conditions. 
This equation essentially neglects rotational inertia of beam elementary elements $\delta z$, and {\it all shearing forces}.

When dealing with small displacements, Eq. (\ref{EB}) is solved by a {\it linear} superposition of eigenmodes $f_n( z,t )$:
\begin{equation}
f_n( z,t ) = x_n(t) \psi_n(z) , \label{fn}
\end{equation}
with $\psi_n(z)$ the {\it mode shape} of mode $n$ (no units), and corresponding mode resonance frequency $\omega_n$. $x_n(t)$ is the time-dependent motion associated with the mode; by means of a Rotating-Frame Transform, it writes $a_{n}(t) \cos (\omega t+ \phi)$ with $a_{n}(t)$ a slow varying amplitude variable, nonzero only for $\omega \approx \omega_n$ (resonance condition). 
Here, $S_z=S_{z,0}=\sigma_0 A_z$ the initially stored axial load in the structure (from uniaxial stress $\sigma_0$).
With this convention, $S_z$ is negative for a {\it tensile} stored stress.
Note that the quantitative value of $x_n(t)$ depends on the normalization choice of $\psi_n(z)$; in this paper we will always normalize modal functions to the maximum displacement amplitude, such that at this abscissa $z_{n}$ one gets $\psi_n(z_{n})=1$. \\

The stretching of the beam writes $S_z=S_{z,0}+\Delta S$ with $\left| \Delta S \right|= E_z A_z \, \Delta L/L$ and $\Delta L$ the extension \cite{crosslifshitzbook}:
\begin{equation}
\Delta L = \frac{1}{2} \int_0^L \!\! \left( \frac{\partial f( z,t ) }{\partial z}\right)^{\!2} dz, \label{Dl}
\end{equation}
expanded at lowest order in $f$. 
Note that from Eq. (\ref{fn}) for a single mode, this expression is quadratic in motion amplitude $x_n(t)$, thus a simple Rotating-Wave Approximation leads to an extension $\Delta L \propto 
a_{n}^2$ (the slow variable): the nonlinear stretching is essentially a {\it static} effect, which is why there is no time-delay in the relationship between $\Delta S$ and $\Delta L$. 
For a superposition of modes, a similar {\it quadratic nonlinear coupling between them} is obtained (see e.g. Ref. \cite{kunalNJP}). \\

		\begin{figure}[t!]
		\centering
		\vspace*{-1.5cm}
	\includegraphics[width=14cm]{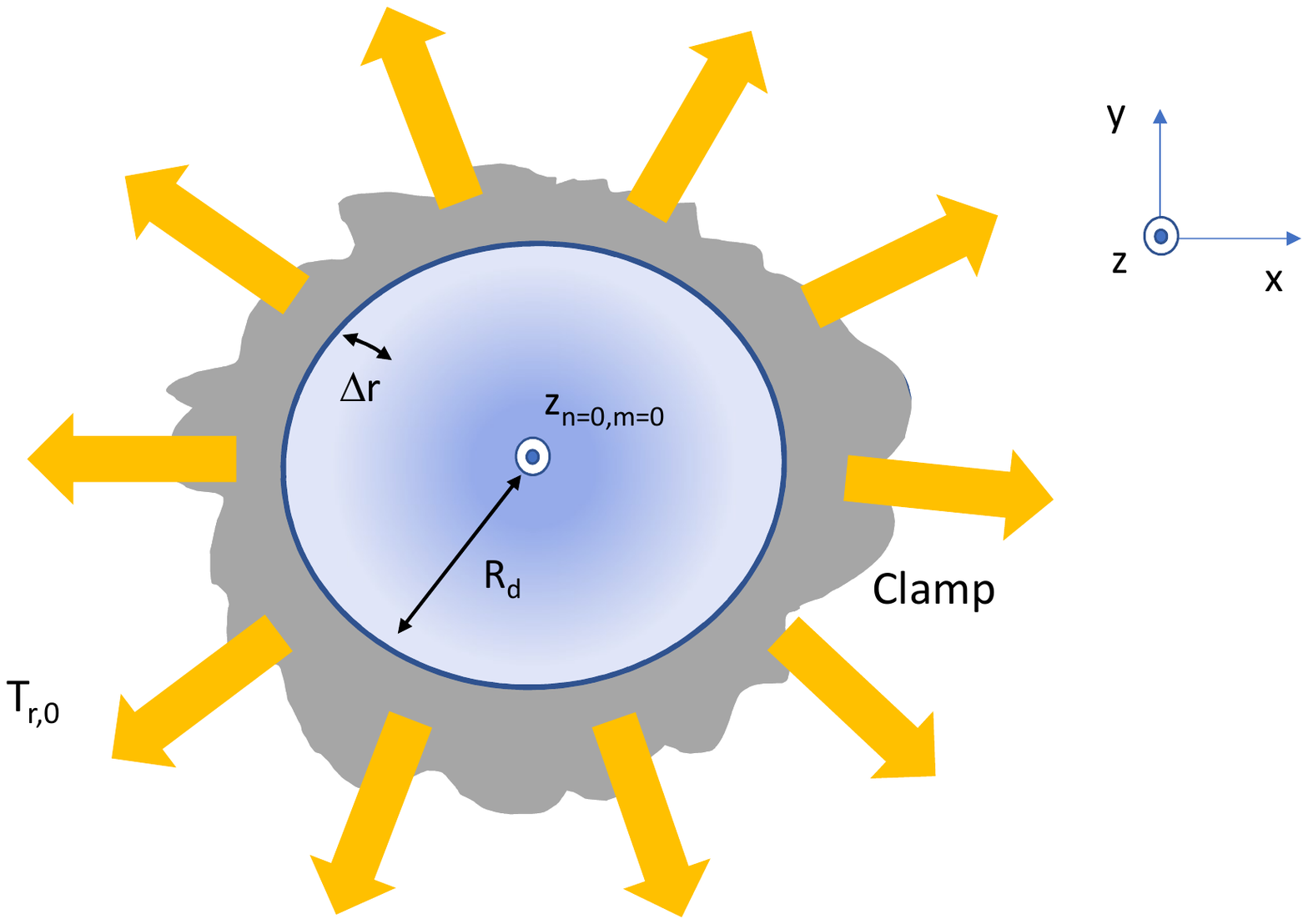}
	\vspace*{-2cm}
			\caption{
			(Color online) Schematic of a drum device, 
		   in its fundamental flexure ($\left\{n=0,m=0\right\}$ mode). The biaxial force $2 \pi R_d \,T_{r,0}$ 
			is here tensile.}
			\label{fig_drum}
		\end{figure}

The basic nonlinear modeling consists then in re-injecting Eq. (\ref{Dl}) into Eq. (\ref{EB}), and {\it neglecting any other alterations due to the large motion amplitude} (see discussion below).		
For a single mode $f \rightarrow f_n$, the projection of Eq. (\ref{EB}) onto it (i.e. multiplying the equation by $\psi_n$ and integrating over the beam length) leads to the definition of {\it modal parameters}:
\begin{eqnarray} 
   \! \! m_n & \! \!\!  =&  \!\! \!\rho A_z  \! \! \int_0^L \!\! \left[ \psi_n(z) \right]^2 dz,  \label{mass} \\
  \! \! k_n &\! \!\!   = &  \!\! \! E_z I_z \! \! \int_0^L \!\! \left[\frac{d^2 \psi_n(z)}{d z^2} \right]^2 \! \!\! dz -  S_{z,0}  \! \! \int_0^L \!\! \left[\frac{d \psi_n(z)}{d z} \right]^2 \! \!\! dz ,  \label{spring} \\
\tilde{k}_n &\! \!\!   = &  \!\! \! \frac{E_z A_z}{2 \, L} \! \left( \int_0^L \!\! \left[\frac{d \psi_n(z)}{d z} \right]^2 \! \!\! dz \right)^{\!\!2} \!, \label{duff}
\end{eqnarray}
with $m_n$ the mode mass, $k_n$ the mode spring constant and $\tilde{k}_n$ the Duffing nonlinear parameter.
The resonance frequency verifies $\omega_n=\sqrt{k_n/m_n}$.
Including in Eq. (\ref{EB}) a damping and a drive term is straightforward \cite{crosslifshitzbook}.
The obtained equation of motion for $x_n$ is then the one of a harmonic oscillator plus a {\it purely cubic nonlinear restoring term}  $+\tilde{k}_n \, x_n(t)^3$.
$\tilde{k}_n$ is always positive, because of stretching (the mode ``hardens'');
in the steady-state ($a_{n}=$ constant), the resonant response measured while sweeping the drive frequency upwards will be {\it pulled up}, with the frequency at maximum amplitude $a_{n}^{max}$ given by $\omega_n^{res} = \omega_n + \beta_n \, (a_{n}^{max})^2$ with $\beta_n=\frac{3}{8} \omega_n \frac{\tilde{k}_n}{k_n}$ \cite{LLMeca,crosslifshitzbook,PRBEddy}.
The free-decay solution can also be analytically produced \cite{PRBEddy}.\\

Beyond the agreement with experiments already mentioned, a discussion on the genesis and validity of this theory is in order.
A thorough discussion of the historical developments can be found in e.g. Ref. \cite{mohammad,nayfehbook}.
The first attempt to model the stretching is due to Woinowsky-Krieger \cite{Woinowsky}. He considered hinged-hinged bars, and restricted his analysis to the simple approximation $\psi_n(z)= \sin (n \pi \, z/L)$, $n>0$ for the mode shapes. Burgreen \cite{Burgreen} considered the same situation for $n=1$ only, but extended it to the case where a compressive axial load is imposed ($S_{z,0}>0$ here). Eisley \cite{Eisley} proposed also a solution for the first mode of clamped-clamped beams, assuming $\psi_1(z)= 1-\cos (2 \pi z/L)$. In all of these studies, nonlinear effects stemming from {\it inertia} and {\it curvature} were neglected; their main achievement was to produce an analytic solution for the Duffing equation (written for $x_n[t]$) in terms of {\it Jacobi Elliptic} functions \cite{Woinowsky,Burgreen,Eisley}. 
The modeling has then been adapted by Yurke et al. \cite{bush}, defining modal parameters as a function of linear mode shapes $\psi_n(z)$ without a sinewave ansatz.
This is the procedure we reproduced above
; solving the Duffing equation for $x_n(t)$ is an extra step that we do not discuss and can be found in e.g. Refs. \cite{crosslifshitzbook,PRBEddy,LLMeca,nayfehbook}. 

Inertia and curvature nonlinearities at large deflections have been studied by Crespo da Silva and Glynn, first for a clamped-free configuration \cite{crespo1,crespo2} and then for a clamped-sliding one \cite{crespo3}.
It turns out that the obtained dynamics equation are {\it of same order} as the ones obtained for pure stretching (these Refs. extend the problem up to order 3 in $x_n$): the result is thus a similar Duffing-like behavior, and there is no {\it a priori reason} to neglect these terms in the stretching theory. 
Indeed, in a later series of articles, Crespo da Silva considered both extensional and curvature-inertia nonlinearities \cite{crespo4,crespo5}. 
The trial functions used for the mode shapes were here the linear solutions $\psi_n(z)$, the approach re-used later on since Ref. \cite{bush}.
His analysis demonstrated that extensional coefficients in the dynamics equation {\it are dominant} compared to the others \cite{crespo5};
this then justifies not to take the latter into account in Eq. (\ref{duff}).

However, the accuracy of the Euler-Bernoulli nonlinear modeling itself remains questionable. Considering inextensional beams, only inertia and curvature nonlinear terms exist. For macroscopic cantilevers, Anderson et al. \cite{anderson} showed that the first mode displays a hardening nonlinearity, while the second mode displays softening. But more recent experiments using nano-mechanical devices demonstrated that for the first mode, experiments {\it do not} match theory: the measured Duffing coefficient is very small, with even a sign change depending on aspect ratio \cite{betaRoukes}.
To date, this has not been explained to our knowledge.
Finally, one approximation which we did not question so far is the use of the {\it linear mode shape as trial function}. While this is obviously more accurate than a simple sinewave (valid only in specific cases), it is not the exact solution of the nonlinear equation. Beyond approximate models obtained e.g. from the {\it method of multiple scales} \cite{nayfehbook}, an expansion of it can be written as $f_n(z,t)=x_n(t) \psi_n(z) + \sum_{k>1} x_n(t)^k \delta \psi_n^{(k)}(z) $ with $\delta \psi_n^{(k)}(z)$ corrective functions matching the boundary conditions, and verifying $\delta \psi_n^{(k)}(z_n)=0$ (such that $x_n$ remains defined as maximum amplitude deflection).
It is obvious that injecting this expression in Eq. (\ref{EB}), the $\delta \psi_n^{(k)}(z)$ {\it do generate terms} that impact the nonlinear coefficients weighting $x_n^2, x_n^3$ in the dynamics equation. 
Considering the success of the basic modeling for doubly-clamped beams, we have to assume that at least in this configuration the $\delta \psi_n^{(k)}(z)$ contributions remain numerically small; but to our knowledge this has not been demonstrated analytically.

The pragmatic point of the present paper is thus to adapt the very same reasoning applied to doubly-clamped beams to the case of circular drum resonators.
We shall not question the theoretical limits mentioned above, but will compare our result to both theory and experiments from the literature.

\begin{figure}[t!]
	\centering		 
	\vspace*{-0cm}
			 \includegraphics[width=7cm]{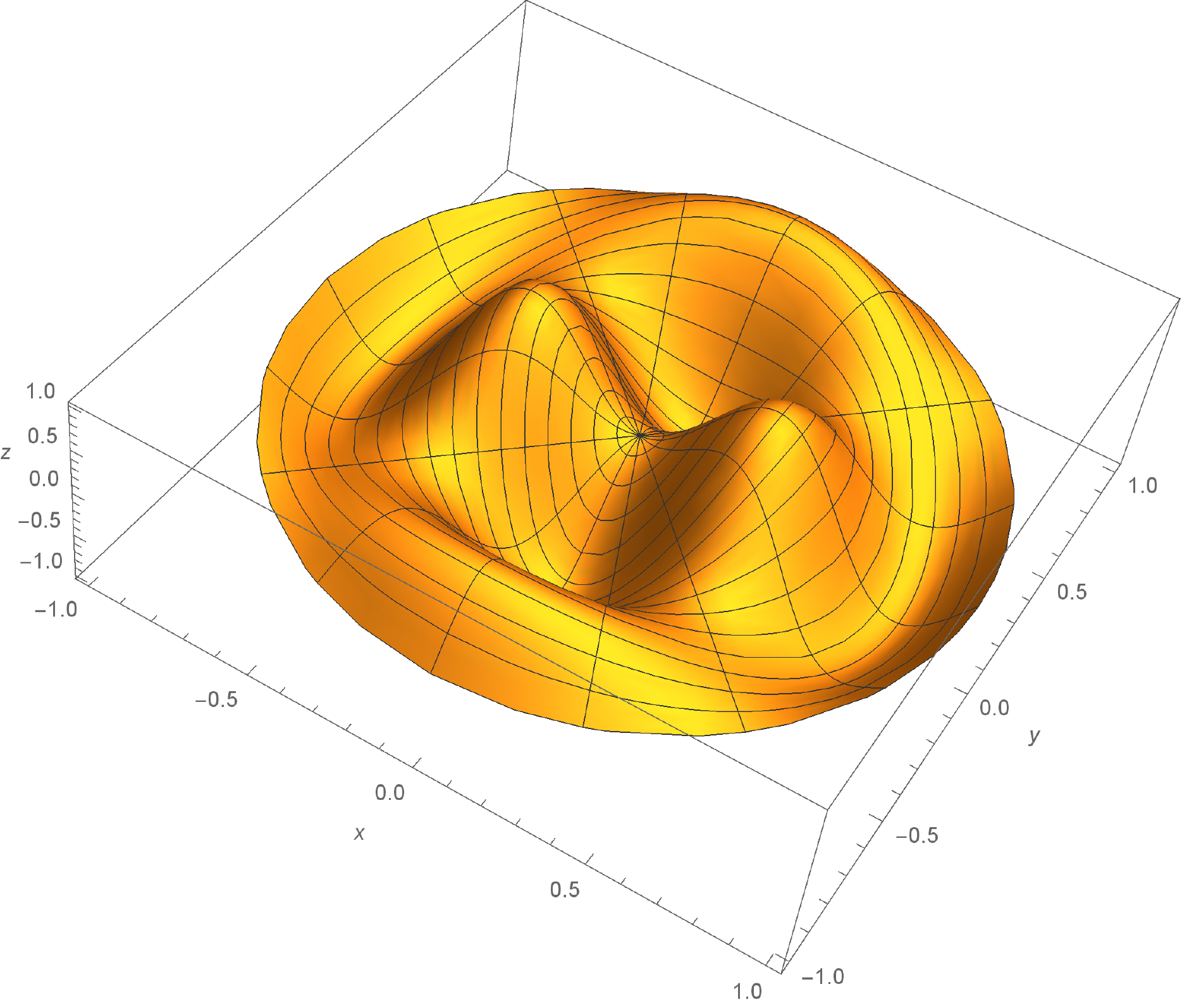}			 		 \includegraphics[width=7cm]{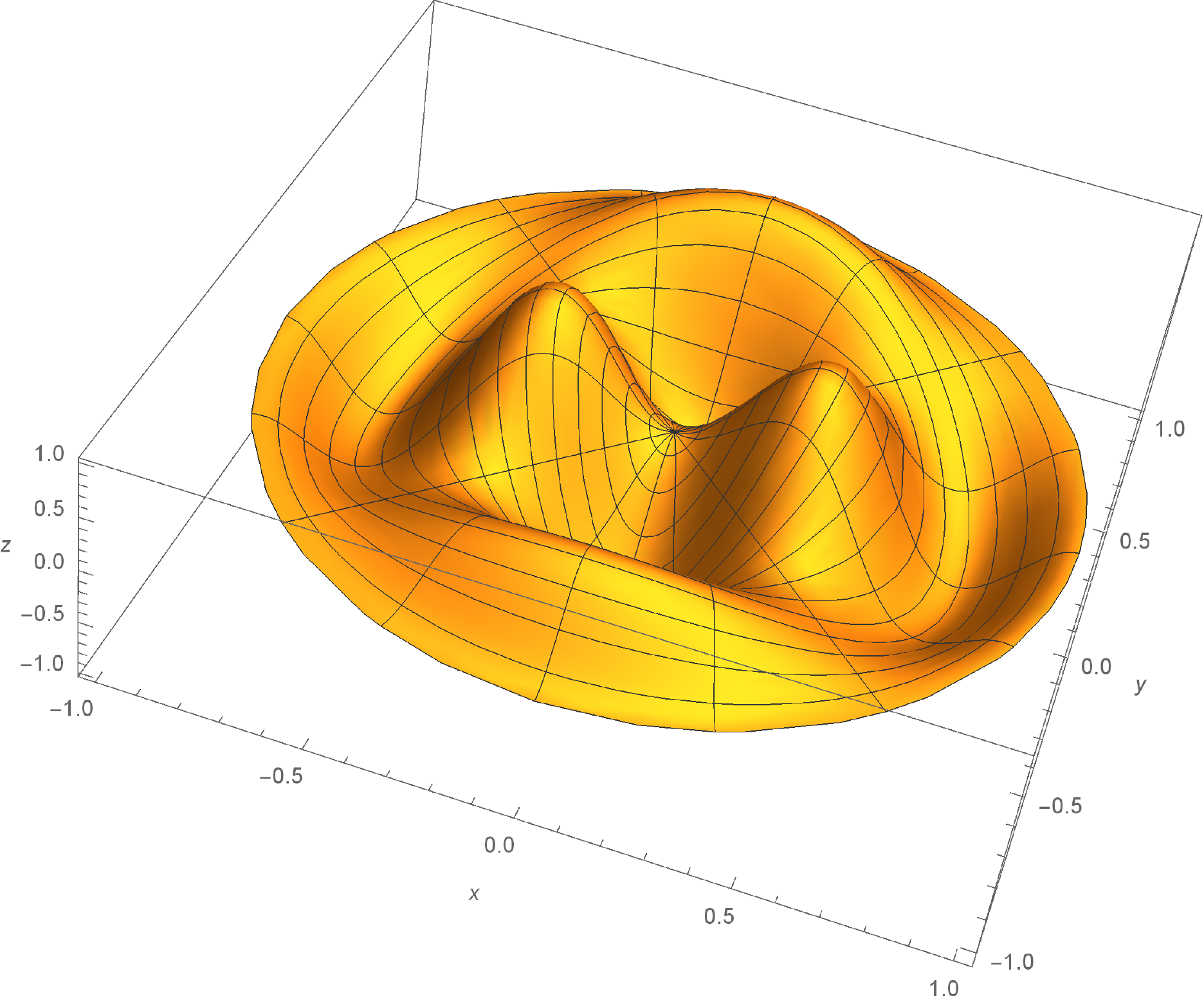}
			\vspace*{-0cm}
			\caption{ (Color online) Calculated mode shape $\psi_{n,m}(r,\theta)$ for mode  $\left\{n=2,m=1\right\}$ (radius $R_d=1$). Top: high-stress limit. Bottom: low-stress limit. Both are very similar in topography. }
			\label{fig_modes}
		\end{figure}

\section{Formulation of the problem}
\label{problem}
	
We now develop the same ideas for the case of a 2D circular structure, see Fig. \ref{fig_drum}. 
We first remind the reader about the conventional {\it linear} theory \cite{schmidbook}.
The generic formalism applying to thin drums [obtained within the same reasoning as Eq. (\ref{EB})] is the Kirchhoff-Love equation:
\begin{equation}
 D_{r} \, \Delta^2 f (r,\theta,t) + T_{r,0}\, \Delta f (r,\theta,t)  =  - \rho h \frac{\partial^2  f (r,\theta,t)}{\partial t^2} , \label{KL}
\end{equation}
with $\Delta \cdots=\frac{1}{r} \frac{\partial}{\partial r} (r \frac{\partial \cdots}{\partial r})+ \frac{1}{r^2}\frac{\partial^2 \cdots}{\partial \theta^2}$ the Laplacian operator (here in polar coordinates), $D_{r} = \frac{1}{12} E_{r} h^3/(1-\nu_{r}^2) $ the flexural rigidity in the plane of the drum ($\nu_{r}$ being Poisson's ratio), $2 \pi R_d \, T_{r,0} =2 \pi R_d  h\, \sigma_0$ the tension within the drum, $h$ its thickness and $R_d$ its radius.
We assume materials properties $E_r, \nu_r, \rho, \sigma_0$ and thickness $h$  to be homogeneous and isotropic over the device; in Eq. (\ref{KL}), the $T_{r,0}$ term resulting from the biaxial stress $\sigma_0$ is taken negative for {\it tensile} load.

		\begin{figure}[t!]
		\centering
		\vspace*{-0.5cm}
		\hspace*{-2cm}
	\includegraphics[width=12cm]{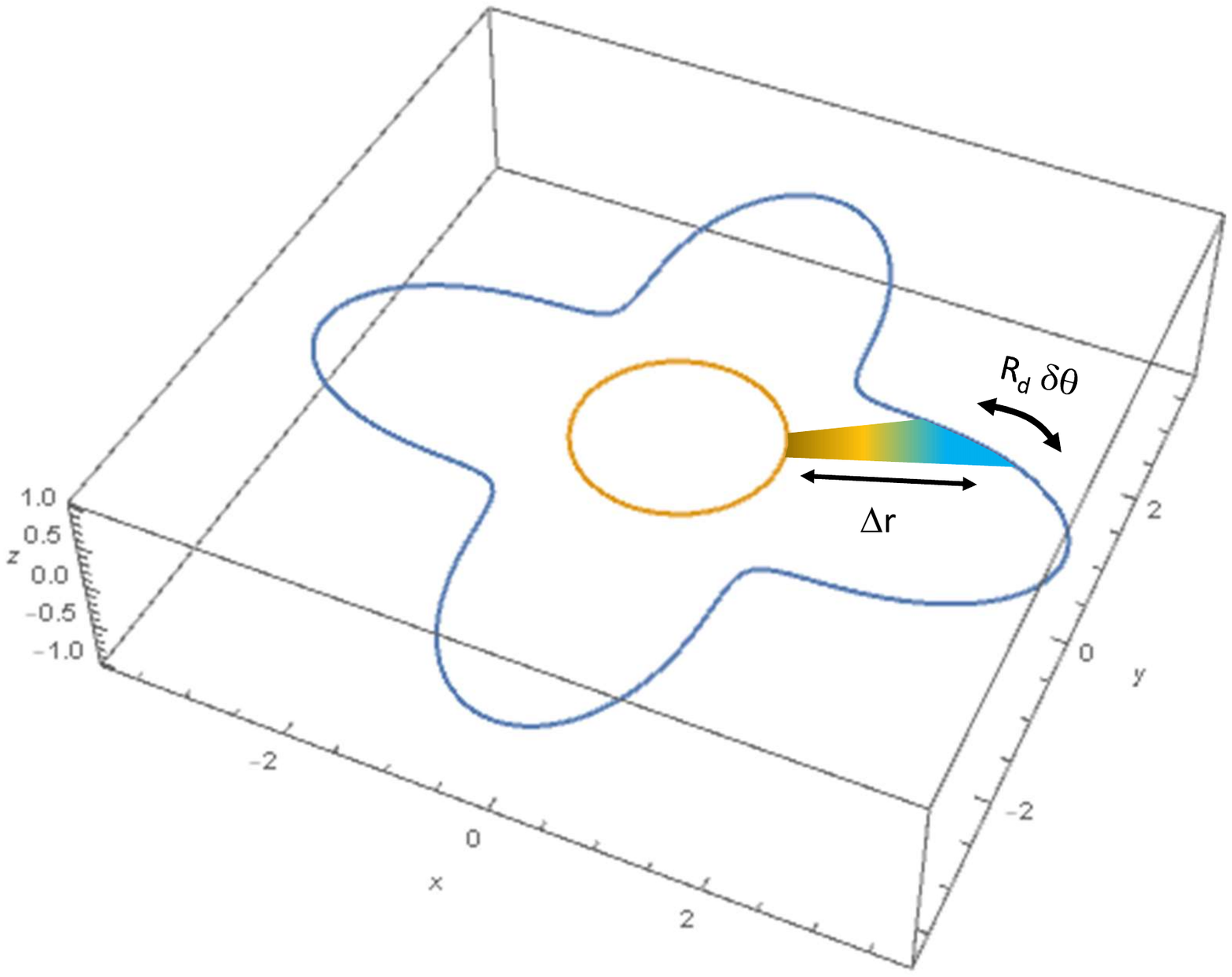}
	\vspace*{-1.cm}
			\caption{
			(Color online) Normalized stretching function $\epsilon$ plotted for mode $\left\{n=2,m=1\right\}$ in high-stress limit (with $z_{n,m}/R_d=1$). The stretched surface area $R_d \delta \theta \Delta r$ is indicated on the graph (central circle represents the drum). }
			\label{fig_stretch}
		\end{figure}		
	
In the limit of small displacements, we write:
\begin{equation}
f_{n,m} (r,\theta,t) = z_{n,m}(t) \psi_{n,m}(r,\theta) , \label{func2D}
\end{equation}
with $\psi_{n,m}(r,\theta)=\phi_{n,m}(r) \,\cos \left(n \,\theta \right)$ the mode shapes and $z_{n,m}(t)$ the motion amplitude; now two indexes are necessary to label all 2D flexural modes of the structure.
Two simple limits are considered in this paper: the high-stress case (membranes, $D_{r}=0$ with $T_{r,0}<0$ here), and the low-stress one (plates, $T_{r,0}=0$). Using the boundary conditions, the solutions write:
\begin{eqnarray}
& & \phi_{n,m}(r) =     \label{bessels}\\
& & \!\!\!\!\!\!\!\!\!\! \frac{\mbox{BesselJ}_n \! \left(\frac{ \lambda_{n,m} r}{R_d} \right)}{\mbox{BesselJ}_n \! \left(\frac{ \lambda_{n,m} r_{n,m}}{R_d} \right)} , \nonumber \\
& &\mbox{or} \nonumber \\
& & \!\!\!\!\!\!\!\!\!\! \frac{\mbox{BesselI}_n \! \left(\frac{ \lambda_{n,m}   r}{R_d} \right)-\frac{\mbox{BesselI}_n \!\left( \lambda_{n,m} \right)}{\mbox{BesselJ}_n \!\left( \lambda_{n,m} \right)} \mbox{BesselJ}_n \!\left( \frac{\lambda_{n,m}   r}{R_d} \right)}{\mbox{BesselI}_n \!\left(\frac{ \lambda_{n,m}   r_{n,m}}{R_d} \right)-\frac{\mbox{BesselI}_n \!\left( \lambda_{n,m} \right)}{\mbox{BesselJ}_n \!\left( \lambda_{n,m} \right)} \mbox{BesselJ}_n \!\left(  \frac{ \lambda_{n,m} r_{n,m}}{R_d} \right)}  ,\nonumber 
\end{eqnarray}
for high-stress and low-stress  respectively. $\lambda_{n,m}$ is the mode parameter and $r_{n,m}$ the radial position of the maximum amplitude (occurring for given angles $\theta$ when $n \neq 0$).
We give the first modes $\lambda_{n,m}$ and $r_{n,m}$ in Tab. \ref{tab1} (Appendix \ref{app1}); the mode $\left\{n=2,m=1\right\}$ is displayed as an example in Fig. \ref{fig_modes} for the two limits (top: high-stress, bottom: low-stress).
	
The stretching in 2D is {\it a change of surface area per unit angle}. This writes mathematically:
\begin{eqnarray}
&&\frac{\delta S}{\delta \theta} =  \label{dS} \\
&&\frac{1}{2} \int_0^{R_d} \!\! \left[ \left( \frac{\partial f( r, \theta ,t ) }{\partial r} \right)^{\!2} + \frac{1}{r^2} \left( \frac{\partial f( r, \theta ,t ) }{\partial \theta} \right)^{\!2} \right] r dr, \nonumber 
\end{eqnarray}
at lowest order in $f$.
Geometrically, this quantity is directly linked to the {\it radial} strain $\epsilon=\Delta r/R_d$ experienced by the drum at its edge: $\delta S = R_d \delta \theta \, \Delta r$, i.e. $\frac{\delta S (\theta,t)}{\delta \theta} = R_d^2 \, \epsilon(\theta,t)$ [see Fig. \ref{fig_stretch}].
Injecting the mode shape Eq. (\ref{func2D}) into Eq. (\ref{dS}), one obtains:
\begin{eqnarray}
&& \epsilon(\theta,t) = \left( \frac{z_{n,m}(t)}{R_d}\right)^2 \times \label{epsilon}\\
&& \left[ \frac{ C_{n,m}^{(1)}+C_{n,m}^{(2)} }{2} + \frac{ C_{n,m}^{(1)}-C_{n,m}^{(2)} }{2} \cos (2n\, \theta)   \right]   , \nonumber
\end{eqnarray}
where we have defined (constants with no dimensions):
\begin{eqnarray}
C_{n,m}^{(1)}&=& \frac{1}{2} \int_0^{R_d} \!\! \left( \frac{d \phi_{n,m}(r) }{d r} \right)^2 r dr , \\
C_{0,m}^{(2)}&=& C_{0,m}^{(1)} , \\
C_{n,m}^{(2)}&=& \frac{1}{2} \int_0^{R_d} \! \frac{n^2}{r^2} \, \phi_{n,m}(r)^2 \, r dr  \,\, \, \mbox{for n$\neq0$}. \nonumber
\end{eqnarray}
The $C_{n,m}^{(1,2)}$ constants of the first modes are given in Tab. \ref{tab4}, Appendix \ref{modeparams}.
We omit indexes $n,m$ in the labeling of $\epsilon$ for simplicity. The function Eq. (\ref{epsilon}) is plotted in Fig. \ref{fig_stretch} for mode $\left\{n=2,m=1\right\}$ in the high-stress limit.

For $n=0$, the problem is isotropic and the solution rather straightforward. However for $n \neq 0$, the stress within the drum {\it has an extra angle-dependent component} $\cos (2n\, \theta)$. Eq. (\ref{KL}) has thus to be modified to:

\begin{eqnarray}
 & &\!\!\!\!\!\!\!\!\!\!\!\! D_{r} \, \Delta^2 f  \label{KLmodif} \\
& + & \int_{-h/2}^{+h/2} \frac{1}{r} \frac{\partial}{\partial r}\left(\!\sigma_r  \, r \frac{\partial f}{\partial r} \right)+  \frac{1}{r^2} \frac{\partial}{\partial \theta}\left(\!\sigma_\theta  \frac{\partial f}{\partial \theta} \right) dz \nonumber \\ 
& = & - \rho h \frac{\partial^2  f }{\partial t^2} ,\nonumber 
\end{eqnarray}
with $\sigma_r(r,\theta,z,t), \sigma_\theta(r,\theta,z,t)$ the superposition of the initial biaxial stress $\sigma_0$ {\it plus} the elastic response of the drum to the strain $\epsilon$, Eq. (\ref{epsilon}).
These stress components are defined below.
As for beams, we neglect any other nonlinear contribution arising from the large motion amplitude; shear stresses (e.g. $\sigma_{r,\theta}$ component) are not taken into account in Kirchhoff-Love theory (as in Euler-Bernoulli).

\section{Stress field}
\label{stressfield}

The next step is thus to compute the stress field within the device; this is indeed the extra difficulty that arises in 2D.
As for beams, we assume that the stretching is adiabatic, i.e. the stress/strain relation can be treated in a time-independent manner.
The total stress field is the sum of a {\it homogeneous} contribution, plus the response to the {\it angle-dependent} stretching.
The former is straightforward (e.g. Appendix \ref{stresssol}):
\begin{eqnarray}
\sigma_r^{hom.}      & = & \sigma_0 - E_r \frac{1}{1-\nu_r} \epsilon^{hom.}, \label{hom1} \\
\sigma_\theta^{hom.} & = & \sigma_0 - E_r \frac{1}{1-\nu_r} \epsilon^{hom.} , \\
\sigma_z^{hom.}      & = &  0 ,\label{hom3} 
\end{eqnarray}
with all shears equal to zero $\sigma_{r,z}=\sigma_{r,\theta}=\sigma_{\theta,z}=0$. 
The $-$ sign above comes from our stress convention. 
In the problem at stake, from Eq. (\ref{epsilon}) we have $\epsilon^{hom.}=\left( \frac{z_{n,m}}{R_d}\right)^2 \left[ \frac{ C_{n,m}^{(1)}+C_{n,m}^{(2)} }{2} \right]$.
This stress field component remains {\it biaxial}. \\

To compute the angle-dependent term, we start with an {\it ansatz} for the associated displacement field $\left\{ u_r,u_\theta,u_z \right\}$:
\begin{eqnarray}
u_r      & = & R_d \, f_r (r,z)      \, \epsilon^{angl.}(\theta) ,  \label{equaur}\\
u_\theta & = & R_d \, f_\theta (r,z) \, \frac{d\epsilon^{angl.}(\theta)}{d \theta} , \\
u_z      & = & h   \, f_z (r,z)      \, \epsilon^{angl.}(\theta) ,  \label{equauz}
\end{eqnarray}
with $\epsilon^{angl.}=\left( \frac{z_{n,m}}{R_d}\right)^2 \left[ \frac{ C_{n,m}^{(1)}-C_{n,m}^{(2)} }{2} \cos (2n\, \theta)  \right]$.
These expressions are then injected in the well-known {\it equilibrium equations} of elasticity theory (see e.g. \cite{clelandbook}), neglecting inertial terms;
these are given for the interested reader in Appendix \ref{stresssol}.

Introducing reduced variables $\tilde{r}=r/R_d$ and $\tilde{z}=z/h$, one can show that the displacement functions have to be written, at lowest order in $h/R_d \ll 1$ (thin structure):
\begin{eqnarray}
f_r (\tilde{r},\tilde{z})      & = &  c_r(\tilde{r})     \, \left|\tilde{z}\right| + b_r(\tilde{r}) + a_r(\tilde{r})          \, \left(\tilde{z}\right)^2 \left[\frac{h}{R_d}\right]^2 \!\!\!  , \label{fr} \\
f_\theta (\tilde{r},\tilde{z}) & = &  c_\theta(\tilde{r}) \, \left|\tilde{z}\right| + b_\theta(\tilde{r}) + a_\theta(\tilde{r})\, \left(\tilde{z}\right)^2 \left[\frac{h}{R_d}\right]^2 \!\!\!  , \\
f_z (\tilde{r},\tilde{z})      & = &   c_z(\tilde{r})     \, \left|\tilde{z}\right| + b_z(\tilde{r}) + a_z(\tilde{r})          \, \left(\tilde{z}\right)^2 \left[\frac{h}{R_d}\right]^2  \nonumber \\
& & \!\!\!\!\!\!\!\!\!\!\!\!\!\!\!\!\!\!\!\!\!\!\!\!\!\!\!\!\!\!\!\!\!\!\!\!\! -\frac{1}{4(1-\nu_r)} \left( \frac{c_r(\tilde{r})- (2 n)^2 \, c_\theta(\tilde{r}) + \tilde{r} \, \frac{d c_r(\tilde{r})}{d\tilde{r}} }{\tilde{r}} \right) \! \left(\tilde{z}\right)^2  \!\!.  \label{fz}
\end{eqnarray} 

For the nine (adimensional) functions $a_X,b_X,c_X$ ($X=r,\theta,z$) of the $\tilde{r}$-variable, we then chose the following  {\it ansatz}:
\begin{eqnarray}
b_r(\tilde{r} )       & = &  b_{r,0}      \,  \tilde{r}^\alpha  , \\
b_\theta(\tilde{r} )  & = &  b_{\theta,0} \,  \tilde{r}^\alpha  , \label{eq1}
\end{eqnarray}
\begin{eqnarray}
c_r(\tilde{r} )       & = &  c_{r,0}      \,  \tilde{r}^{\alpha}  , \\
c_\theta(\tilde{r} )  & = &  c_{\theta,0} \,  \tilde{r}^{\alpha}  , 
\end{eqnarray}
\begin{eqnarray}
b_z(\tilde{r} )       & = &  b_{z,0} \, \tilde{r}^{\alpha-1}  , \\
c_z(\tilde{r} )       & = &  c_{z,0} \, \tilde{r}^{\alpha-1}  , 
\end{eqnarray}
\begin{eqnarray}
a_r(\tilde{r} )       & = &  a_{r,0}      \, \tilde{r}^{\alpha-2}   , \\
a_\theta(\tilde{r} )  & = &  a_{\theta,0} \, \tilde{r}^{\alpha-2}  , 
\end{eqnarray}
and:
\begin{equation}
a_z(\tilde{r} )        =   a_{z,0}      \, \tilde{r}^{\alpha-3}  , \label{eqn}
\end{equation}
which leads to seven equations linking the above introduced constants. 
Obviously, $\alpha \geq 3$ to guarantee a physical solution.

\begin{figure}[t!]
	\centering		 
	\vspace*{-0cm}
			 \includegraphics[width=7.5cm]{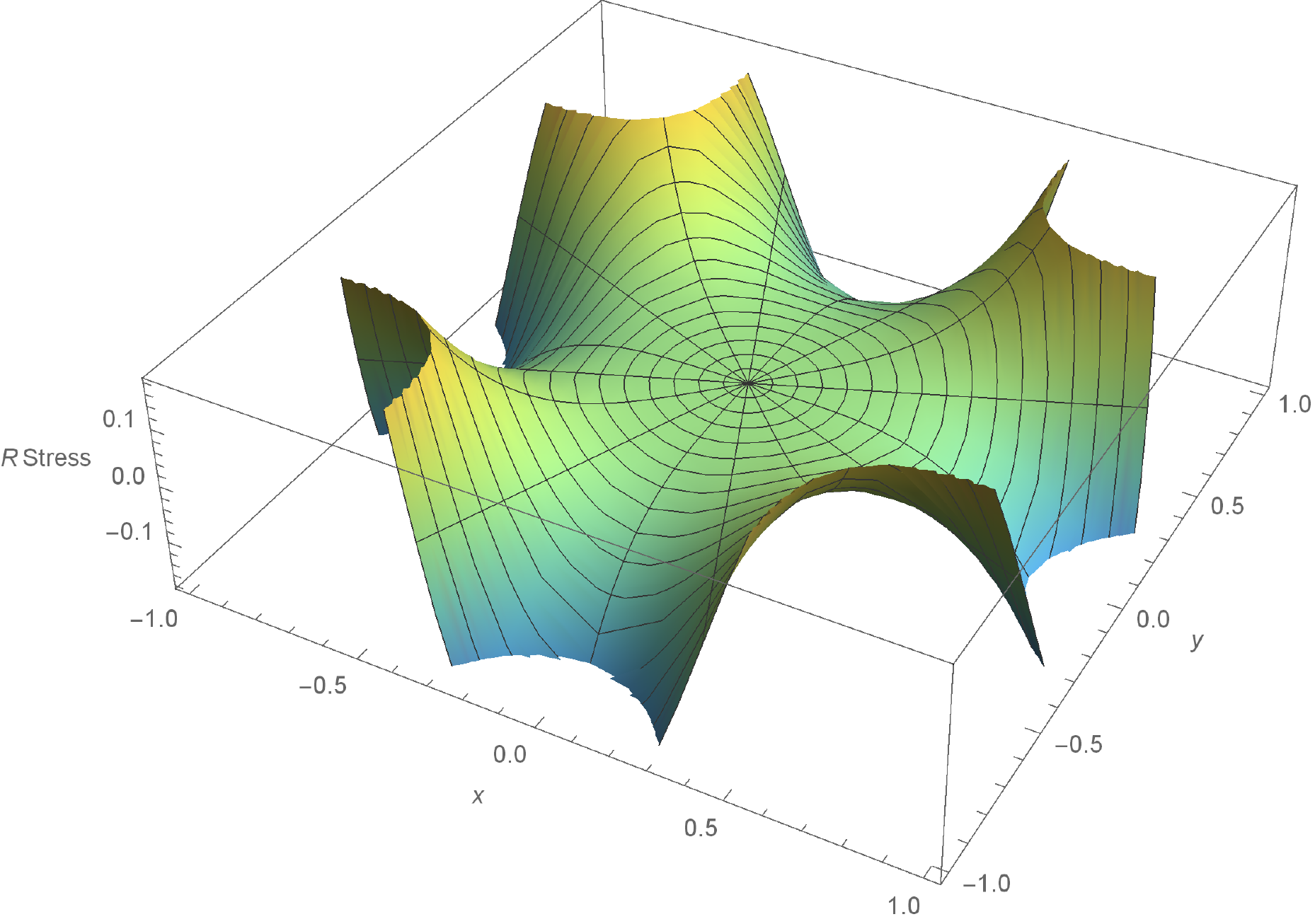}			 		 \includegraphics[width=7.5cm]{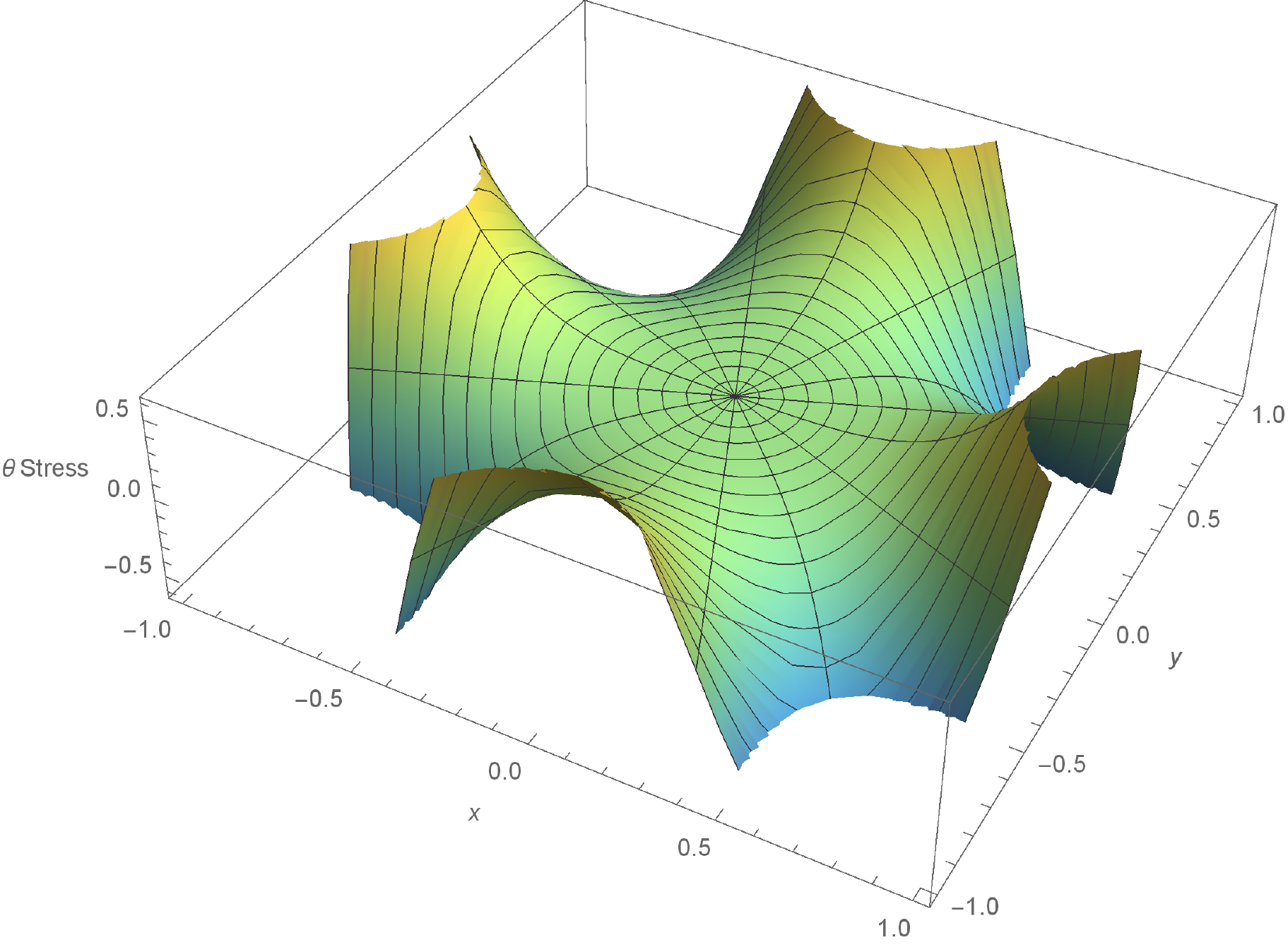}
			\vspace*{-0cm}
			\caption{ (Color online) Stress components $\sigma_r^{angl.}$ (top) and $\sigma_\theta^{angl.}$ (bottom) computed for mode $\left\{n=2,m=1\right\}$ in the high-stress limit. The graph has been normalized to $E_r=1$, $R_d=1$, and $z_{n,m}/R_d=1$ (using $\nu_r=+0.3$).}
			\label{fig_stress}
		\end{figure}
		
Three more equations are obtained from the stress boundary conditions on the surface of the drum: $\sigma_z(r,\theta,z=\pm h/2)=0$, 
$\sigma_{r,z}(r,\theta,z=\pm h/2)=0$ and $\sigma_{\theta,z}(r,\theta,z=\pm h/2)=0$. 
The last relation is obtained from the stretching on the periphery, equating the radial strain $\partial u_r/\partial r$ to $\epsilon^{angl.}$
at $r=R_d$ (see Fig. \ref{fig_stretch}).
Solving the problem under Mathematica$^{\textregistered}$, we list 
the constants appearing in Eqs. (\ref{eq1}-\ref{eqn}) in Tab. \ref{tab2}, Appendix \ref{stresssol} (as a function of $n$ and $\nu_r$). The exponent $\alpha$ is found to be $2 n +1$, reminding $n \neq 0$.

The $\theta$-dependent stress field can finally be calculated. The normal components write, in the limit $h/R_d \approx 0$:
\begin{eqnarray}
\sigma_r^{angl.}      & = & -E_r \, \eta_r^{(n)}(\nu_r)      \left( \frac{r}{R_d} \right)^{\alpha-1} \!\! \epsilon^{angl.}, \label{angl1}\\
\sigma_\theta^{angl.} & = & -E_r \, \eta_\theta^{(n)}(\nu_r) \left( \frac{r}{R_d} \right)^{\alpha-1} \!\! \epsilon^{angl.}, \label{angl2}\\
\sigma_z^{angl.} & = & 0. \label{angl3}
\end{eqnarray}
The functions $\eta_r^{(n)}(\nu_r)$ and $\eta_\theta^{(n)}(\nu_r)$ with $n\neq0$ are defined by:
\begin{eqnarray}
 \eta_r^{(n\neq0)}(\nu_r)           & =&          \frac{1+2n -2(1+n) \nu_r}{(1+2n)(1+\nu_r)}                  , \label{etas1} \\    
 \eta_\theta^{(n\neq0)}(\nu_r)      & =&          -\frac{3+4n}{(1+2n)(1+\nu_r)}                  . \label{etas2}
\end{eqnarray}
The only nonzero shear stress is $\sigma_{r,\theta}$ (see Appendix \ref{stresssol}). 
It shall be neglected in this modified Kirchhoff-Love theory, as already stated. As an example, the computed (normalized) stress components are displayed in Fig. \ref{fig_stress} for mode $\left\{n=2,m=1\right\}$, in the high-stress limit.

Angle-dependent terms Eqs. (\ref{angl1}-\ref{angl3}) and homogeneous terms Eqs. (\ref{hom1} -\ref{hom3}) can be rewritten in a compact form:
\begin{eqnarray}
\!\!\!\!\!\!\!\!\!\!\!\!\!\!\! \sigma_r      &\!\!\! = &\!\!\! \sigma_0 - E_r \!\! \left[ \eta_r^{(0)}(\nu_r)   \,   \epsilon^{hom.} + \eta_r^{(n)}(\nu_r)       \left( \frac{r}{R_d} \right)^{2 n}  \!\!\!\! \epsilon^{angl.}\right] \!,  \\
\!\!\!\!\!\!\!\!\!\!\!\!\!\!\! \sigma_\theta &\!\!\! = &\!\!\! \sigma_0 - E_r \!\! \left[ \eta_\theta^{(0)}(\nu_r) \, \epsilon^{hom.} + \eta_\theta^{(n)}(\nu_r)  \left( \frac{r}{R_d} \right)^{2 n} \!\!\!\! \epsilon^{angl.}\right] \!,  \\
\!\!\!\!\!\!\!\!\!\!\!\!\!\!\! \sigma_z      &\!\!\! = &\!\!\! 0  ,   
\end{eqnarray}
provided we define $\eta_r^{(0)}(\nu_r) =\eta_\theta^{(0)}(\nu_r)=\eta^{(0)}(\nu_r)=1/(1-\nu_r)$.
The stress is still {\it planar}, and independent of $z$, but $\sigma_r  \neq \sigma_\theta$ and is neither homogeneous nor isotropic.
Injecting these in Eq. (\ref{KLmodif}), we can now solve the problem at hand. 

\section{Mode parameters}
\label{soluce}
	
Having found the stress field, we can now project	Eq. (\ref{KLmodif}) on a given mode $\left\{n,m\right\}$.
We thus define modal parameters:
\begin{eqnarray}
{\cal M}_{n,m} & = & \rho h \int_0^{2 \pi} \!\! \int_0^{R_d} \! \left[\psi_{n,m}(r,\theta) \right]^2  rdr d\theta ,\\
{\cal K}_{m,n} & = & D_r \int_0^{2 \pi} \!\! \int_0^{R_d}  \! \left[ \psi_{n,m}(r,\theta) \Delta^2 \psi_{n,m}(r,\theta) \right] \, rdr d\theta  \nonumber \\
&\!\!\!\!\!\!\!\!\!\!\!\!\!\!\!\!\!\!\!\!\!\!\!\!\!\!\!\!\!\!\!\! + &\!\!\!\!\!\!\!\!\!\!\!\!\!\!\!  T_{r,0}  \int_0^{2 \pi} \!\! \int_0^{R_d} \left[ \psi_{n,m}(r,\theta) \Delta  \psi_{n,m}(r,\theta) \right] \, rdr d\theta, 
\end{eqnarray}
in a similar fashion to Eqs. (\ref{mass} - \ref{spring}).
The resonance frequencies $\omega_{n,m}=\sqrt{ {\cal K}_{m,n}/{\cal M}_{n,m} }$  reduce to:
\begin{eqnarray}
& & \omega_{n,m}  =   \label{freqref}\\
& & \sqrt{\frac{\left|T_{r,0}\right|}{\rho \, h}} \left(\frac{\lambda_{n,m}}{R_d}\right) \nonumber \\
& & \mbox{or} \nonumber \\
& & \sqrt{\frac{D_r }{\rho \, h}} \left(\frac{\lambda_{n,m}}{R_d}\right)^{\!2} , \nonumber 
\end{eqnarray}
in the limit of high-stress and low-stress devices, respectively. 
We give mass and spring values for the first modes in Tab. \ref{tab3}, Appendix \ref{modeparams}.

Beyond the usual linear coefficients, the Duffing term analogous to Eq. (\ref{duff}) finally writes:
\begin{eqnarray}
\!\!\!\! \tilde{{\cal K}}_{n,m} & = & - \frac{ E_r h}{R_d^2}   \times \label{duffKL} \\
& & \!\!\!\!\!\!\!\!\!\!\!\!\!\!\!\!\!\!\!\!\!\!\!\!\!\!\! \left[ \frac{ C_{n,m}^{(1)}+C_{n,m}^{(2)} }{2}  \, \eta^{(0)}(\nu_r) \int_0^{2 \pi} \!\!\! \int_0^{1} \left[ \psi_{n,m}(\tilde{r},\theta) \Delta  \psi_{n,m}(\tilde{r},\theta) \right] \, \tilde{r}d \tilde{r} d\theta  \right.   \nonumber \\
&\!\!\!\!\!\!\! + &\!\!\!\!\!\!\! \left. \frac{ C_{n,m}^{(1)}-C_{n,m}^{(2)} }{2}  \times \right. \nonumber \\
& & \left. \left( \eta_r^{(n)}(\nu_r) \frac{\pi}{2} \int_0^{1} \frac{\phi_{n,m}(\tilde{r})}{\tilde{r}} \frac{d}{d \tilde{r}} \left(\tilde{r}^{2n+1} \, \frac{d \phi_{n,m}(\tilde{r})}{d \tilde{r}} \right) \tilde{r} d\tilde{r} \right. \right. \nonumber  \\
& + & \left. \left. \eta_\theta^{(n)}(\nu_r)  \frac{n^2 \pi}{2} \int_0^{1} \frac{\tilde{r}^{2n+1} \, \phi_{n,m}(\tilde{r})^2 }{\tilde{r}^3} \tilde{r} d\tilde{r} \right) \right] , \nonumber 
\end{eqnarray}
with the integrals written in normalized units $\tilde{r}=r/R_d$ (no dimensions). 

Numerical values for the integrals defining the coefficients $\tilde{{\cal K}}_{n,m}$ are listed and discussed for the first modes in Appendix \ref{modeparams},
Tabs. \ref{tab5} and \ref{tab5bis}.
For beams, Eq. (\ref{duff}) leads to a scaling of the Duffing parameter $\tilde{k}_n \propto E_z A_z/L^3$.
Similarly here, Eq. (\ref{duffKL}) leads to $\tilde{{\cal K}}_{n,m} \propto E_r (h \, 2 \pi R_d)/R_d^3$; in both cases, the Duffing effect is a {\it stiffening}. 

\section{Discussion}
\label{discuss}

As for beams, there is a tremendous literature on nonlinear plates and membranes.
In Section \ref{beambase}, we reviewed the beam-based modeling in order to clarify the basis of the theory that we adapt here in 2D; for a detailed account of historical developments in the modeling of drums, we direct the interested reader to Refs. \cite{Amabili_book,Pai_book,nayfehbook}.

		\begin{figure}[t!]
		\centering
		\vspace*{-0.3cm}
	\includegraphics[width=11cm]{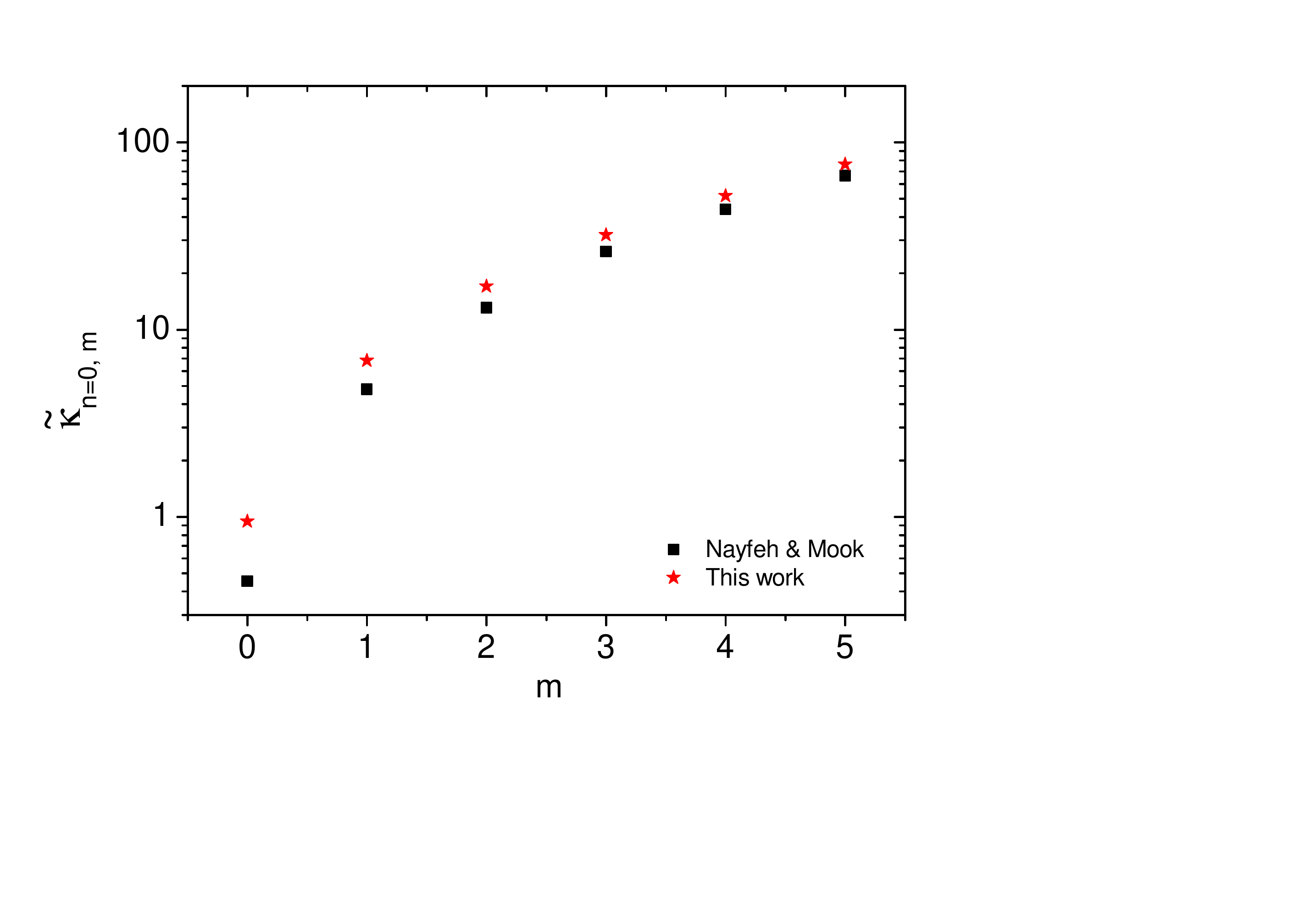}
	\vspace*{-2.3cm}
			\caption{
			(Color online) Comparison between our theory (stars) and Ref. \cite{nayfehbook} (squares) in units of $E_r h/R_d^2$, computed for $\nu=1/3$; see text.}
			\label{fig_compare}
		\end{figure}


Let us however illustrate the theoretical state-of-art with typical results from the field of MEMS and NEMS.
In the last decades, the fast development of micro and nano-mechanics has been an impetus to new theoretical support, especially using modern numerical computation capabilities.
Especially, the use of electrostatic actuation has been directly incorporated in the modeling (e.g. Ref. \cite{kozinsky} for beam-based structures).
For clamped circular plates, 
the conventional approach is to reduce the problem to a system of coupled ordinary differential equations, and ultimately rely on numerical methods for predictions \cite{Vogl1,Vogl2}.
Indeed, numerical integration of nonlinear equations including electrostatic drives has proven to be an extremely efficient tool for fitting experimental data; this is the procedure followed in Ref. \cite{Sajadi} to access values of the Young's moduli in graphene membranes. 

In contrast, our modeling remains at a {\it generic level}, not introducing any specific drive fields: we model only the stretching effect with no hypothesis on the origin of in-built stress. The aim is to produce an analytic expression for the Duffing coefficient.
Besides, all these works deal with axisymmetric modes; Eq. (\ref{duffKL}) applies to {\it any} $\left\{ n,m \right\}$.

From Ref. \cite{nayfehbook}, we reproduce here the analytical modeling of axisymmetric modes.
The nonlinear coefficient $\tilde{{\cal K}}_{n=0,m}^{norm}$ normalized to $E_r h/R_d^2$ is written as:
\begin{equation}
\!\!\!\!\! \tilde{{\cal K}}_{0,m}^{norm}  = \! \sum_{k=1}^{\infty} \frac{ \left[\int_0^1  \left( d \phi_{0,m}[\tilde{r}]/d \tilde{r} \right)^2  \mbox{BesselJ}_1 \! \left( \zeta_k \tilde{r} \right)     d \tilde{r} \right]^2}{\left(\zeta_k^2 -1 + \nu^2 \right) \left[\mbox{BesselJ}_1 \! \left( \zeta_k\right) \right]^2 } ,
\end{equation}
with the $\zeta_k$ parameters tabulated for $\nu=1/3$ therein ($k>0$ integer) \cite{nayfehcorr}.
Comparison with our expression is given in Fig. \ref{fig_compare}, summing up to $k=12$. In the displayed range, this leads to a numerical accuracy better than the size of the symbols. 
The two models converge towards each other for large $m$, with our numerical value above the one of Ref. \cite{nayfehbook}; for $m>1$ the difference is less than about $30~\%$.  \\

Beyond comparison to existing theory, we shall assess the validity of our modeling by comparing it to benchmark measured devices from the literature in their $\left\{n=0,m=0\right\}$ resonance: a MEMS type silicon-nitride membrane in the high-stress limit, a top-down graphene NEMS device (low stress) and finally an aluminum drumhead NEMS that is typically used in quantum electronics experiments.

In Ref. \cite{DuffWeig} the nonlinear behavior of a square-like silicon nitride drum has been studied.
From Fig. 1 (b) of this article, we infer a Duffing parameter normalized to the mode mass of about $\tilde{{\cal K}}_{0,0}/{\cal M}_{0,0} \approx +1.5\, 10^{23}~$m$^{-2}$s$^{-2}$.
This fits the data for weak enough excitations; with larger drives, other nonlinear features kick in \cite{DuffWeig}.
Even though the initial stress $\sigma_0$ stored in the structure is not very high (110$~$MPa), the device is well within the {\it membrane} limit. 
Looking at Fig.  2 (a) from Ref. \cite{DuffWeig} which displays the optically measured pattern of the first mode, it appears that 
it can be accurately approximated by a circular shape of radius $R_d \approx 210~\mu$m.
From the parameters given in the publication (supplementary material, $E_r =240~$GPa, $\rho=3200~$kg/m$^3$, thickness $h=480~$nm), taking a standard Poisson ratio of $\nu_r =+0.3$, we obtain 
$\tilde{{\cal K}}_{0,0}/{\cal M}_{0,0} \approx +1.2\, 10^{23}~$m$^{-2}$s$^{-2}$. 
The corresponding mode resonance frequency calculated is 338$~$kHz, matching also consistently the measured 321$~$kHz.

In Ref. \cite{DuffVdZ}  the nonlinear behavior of a (multilayer) graphene drum has been studied.
The reported stored stress is very low (about 5$~$MPa), and the device is better described in the {\it plate} limit.
From parameters quoted in the publication ($E_r \approx 700~$GPa, $\rho = 600~$kg/m$^3$, $R_d = 2.5~\mu$m, $h = 5~$nm, neglecting the Poisson ratio)
we compute $\tilde{{\cal K}}_{0,0}/{\cal M}_{0,0} \approx +3.3\, 10^{31}~$m$^{-2}$s$^{-2}$ for a resonance frequency of 12.8$~$MHz.
Again, this is in close agreement with measured values of $+2.\, 10^{31}~$m$^{-2}$s$^{-2}$ and 14.5$~$MHz respectively, given in the publication for the zero DC voltage bias limit.

In Ref. \cite{DylanPRX}, the nonlinear dynamics of an opto-mechanical system consisting of an aluminum drumhead device coupled to a microwave cavity has been studied.
The device is about 8.5$~\mu$m in radius and 170$~$nm in thickness (Appendix of Ref. \cite{DylanPRX}), and displays a resonance frequency for the fundamental out-of-plane flexure of 6.8$~$MHz.
The in-built stress $\sigma_0$ is not accurately known, but should be in the range $0 - 60 ~$MPa.
We take for aluminum the bulk values $E_r = 70~$GPa, $\rho = 2700~$kg/m$^3$ and $\nu_r =+0.35$.
The Duffing parameter that is fit onto the mechanical frequency shift (Fig. 4 of Ref. \cite{DylanPRX}) is about $\tilde{{\cal K}}_{0,0}/{\cal M}_{0,0} \approx +7.\, 10^{27}~$m$^{-2}$s$^{-2}$. This matches within $\pm 50~\%$ the theoretical estimate based on our modeling, for high-stress and low-stress limits, with a calculated frequency matching 6.7$~$MHz.

\section{Conclusion}

Following the same methodology as for beams, we present a theory describing the geometrical (stretching) nonlinearity of drum devices. 
The basic hypotheses are to neglect any other nonlinear features apart from the extra tensile stress, to neglect shearing forces, and to treat the stretching as a static effect.
Two limits are considered for numerical estimates: high-stress (membranes) and low-stress (plates), but the mathematical description is written in a generic fashion.
The difficulty lies in the calculation of the stress profile induced in the stretched drum for non-axisymmetric modes; the analytic solution however exists in the limit of a thin structure.

We thus present a {\it simple and fully analytic} modeling of the Duffing nonlinear coefficient $\tilde{{\cal K}}_{n,m}$ of circular plates and membranes. Only the knowledge of the mode shapes is necessary for the calculation of $\tilde{{\cal K}}_{n,m}$, through simple integrals evaluation; 
the first numerical values are given in Appendix.
No hypotheses are made on the drive schemes, neither on the nature of the in-built biaxial stress.
The theory is compared to existing analytics from Ref. \cite{nayfehbook}, and to benchmark experimental data \cite{DuffWeig,DuffVdZ,DylanPRX}. 
In both cases, the agreement is good.

Further comparison with experiments should be done with higher modes, especially non-axisymmetric ones ($n \neq 0$).
Besides, the presented theory can be in principle extended to mode-coupling \cite{kunalNJP,NanoRoukes};
an experimental and theoretical study of this regime would definitely assess the validity of the presented mathematical methods.

\section{Data availability}
Data sharing is not applicable to this article as no new data were created or analyzed in this study.

\begin{acknowledgements}

We acknowledge support from the ERC CoG grant ULT-NEMS No. 647917, StG grant UNIGLASS No. 714692 and the STaRS-MOC project from {\it R\'egion Hauts-de-France}. 
The research leading to these results has received funding from the European Union's Horizon 2020 Research and Innovation Programme, under grant agreement No. 824109, the European Microkelvin Platform (EMP).

\vspace*{0.1cm}
(\dag) Corresponding Author: eddy.collin@neel.cnrs.fr

\end{acknowledgements}


\appendix

\section{Mode parameters}
\label{app1}

In the Table below we give the first modes $\lambda_{n,m}$ and $r_{n,m}$ parameters for both high-stress (H.S.) and low-stress (L.S.) limits.
Inserting these in Eqs. (\ref{bessels}) one can easily compute the corresponding mode shapes (see Fig. \ref{fig_modes} for an example).

\begin{table}[h!]
\begin{center}
\begin{tabular}{|c|c|c|c|c|c|}    \hline
mode     $\left\{n,m\right\}$    &  H.S. $\lambda_{n,m}$        &  H.S. $r_{n,m}$ & & L.S. $\lambda_{n,m}$  &    L.S. $r_{n,m}$  \\    \hline    \hline
 $\left\{0,0\right\}$    &  2.40483       &  0.           & &  3.19622   &  0.                        \\    \hline
 $\left\{0,1\right\}$    &  5.52008       &  0.           & &  6.30644   &  0.                        \\    \hline
 $\left\{1,0\right\}$    &  3.83171       &  0.4805123    & &  4.61090   &  0.4102482                 \\    \hline
 $\left\{1,1\right\}$    &  7.01559       &  0.2624418    & &  7.79927   &  0.2358243                 \\    \hline
 $\left\{0,2\right\}$    &  8.65373       &  0.           & &  9.43950   &  0.                        \\    \hline
 $\left\{2,0\right\}$    &  5.13562       &  0.5947163    & &  5.90568   &  0.5258299                 \\    \hline
 $\left\{1,2\right\}$    &  10.1735       &  0.1809784    & &  10.9581   &  0.1680282                 \\    \hline
 $\left\{2,1\right\}$    &  8.41724       &  0.3628549    & &  9.19688   &  0.3319174                 \\    \hline
 $\cdots$                &  $\cdots$      &  $\cdots$     & &  $\cdots$    &  $\cdots$                \\    \hline
\end{tabular}
\caption{\label{tab1} First mode parameters. Left: high-stress (H.S.), right: low-stress (L.S.). For $n=0$ modes, the maximum amplitude is at the center ($r_{0,m}=0$). }
\end{center}
\end{table}

\section{Stress field solution}
\label{stresssol}

We remind the reader basics of elasticity theory expressed in cylindrical coordinates. 
The strain fields can be written in terms of the displacement fields:
\begin{eqnarray}
\epsilon_r & = & \frac{\partial u_r}{\partial r} , \nonumber \\
\epsilon_\theta & = & \frac{u_r}{r} + \frac{1}{r} \frac{\partial u_\theta}{\partial \theta} , \nonumber  \\
\epsilon_z & = & \frac{\partial u_z}{\partial z} ,\nonumber 
\end{eqnarray}
for the normal components, and:
\begin{eqnarray}
2\epsilon_{r,\theta} & = & \frac{\partial u_\theta}{\partial r}-\frac{u_\theta}{r} + \frac{1}{r} \frac{\partial u_r}{\partial \theta}, \nonumber  \\
2\epsilon_{r,z} & = &  \frac{\partial u_r}{\partial z} + \frac{\partial u_z}{\partial r} , \nonumber  \\
2\epsilon_{\theta,z} & = &  \frac{1}{r} \frac{\partial u_z}{\partial \theta}+ \frac{\partial u_\theta}{\partial z} , \nonumber 
\end{eqnarray}
for the shear strains.

For an isotropic homogeneous Hookean material, we have:
\vspace*{-0.3cm}
\begin{displaymath}
\bordermatrix{ &    \cr
               & \sigma_r \cr
							 & \sigma_\theta \cr
							 & \sigma_z \cr
							 & \sigma_{r,\theta} \cr
							 & \sigma_{r,z} \cr
							 & \sigma_{\theta,z} \cr}  =
							\frac{E_r}{(1+\nu_r)(1-2 \nu_r)} \bordermatrix{ & \cr
							                                                & {\cal H }\cr}						
							\bordermatrix{ &    \cr
               & \epsilon_r \cr
							 & \epsilon_\theta \cr
							 & \epsilon_z \cr
							 & 2\epsilon_{r,\theta} \cr
							 & 2\epsilon_{r,z} \cr
							 & 2\epsilon_{\theta,z} \cr} 
\end{displaymath}
with $( {\cal H} )	=$
\vspace*{-0.3cm}
\begin{displaymath}
									\bordermatrix{  &  & & & & &  \cr
                                  & 1-\nu_r & \nu_r   & \nu_r   & 0 & 0 & 0 \cr
 		  														& \nu_r   & 1-\nu_r & \nu_r   & 0 & 0 & 0 \cr
	 		  													& \nu_r   & \nu_r   & 1-\nu_r & 0 & 0 & 0 \cr
																	& 0       &       0 & 0       & \frac{1-2\nu_r}{2} & 0 & 0 \cr
																	& 0       &       0 & 0       & 0  &\frac{1-2\nu_r}{2} & 0 \cr
																	& 0       &       0 & 0       & 0 & 0 & \frac{1-2\nu_r}{2} \cr} 
\end{displaymath}
for the relationship between stresses $( \sigma )$ and strains $( \epsilon )$.

The {\it equilibrium equations} then write:
\begin{eqnarray}
 \frac{\partial \sigma_r}{\partial r} + \frac{1}{r}\frac{\partial \sigma_{r,\theta}}{\partial \theta}+ \frac{1}{r} \left( \sigma_r-\sigma_\theta\right) +\frac{\partial \sigma_{r,z}}{\partial z} &= & 0, \nonumber \\
\frac{ \partial \sigma_{r,\theta}}{\partial r} + \frac{1}{r}\frac{\partial \sigma_\theta}{\partial \theta} + 2 \frac{\sigma_{r,\theta}}{r} + \frac{ \partial \sigma_{\theta,z}}{\partial z} &= & 0, \nonumber \\
 \frac{ \partial \sigma_{r,z}}{\partial r} + \frac{1}{r}\frac{\partial \sigma_{\theta,z}}{\partial \theta} +\frac{\sigma_{r,z}}{r}  +\frac{\partial \sigma_z}{\partial z}&= & 0, \nonumber
\end{eqnarray}
when neglecting the inertial terms. 

The solution for the {\it homogeneous} stretching component is straightforward.
The well-known displacement field simply writes:
\begin{eqnarray}
f_r (r,z)       &= & r , \nonumber \\
f_\theta (r,z)  &= & 0, \nonumber \\
f_z (r,z)       &= & - \frac{2 \nu_r}{1-\nu_r} z , \nonumber 
\end{eqnarray}
with $u_r=f_r \, \epsilon^{hom.}, u_\theta=0, u_z= f_z \, \epsilon^{hom.}$ by definition.
Then $\epsilon_r=\epsilon_\theta=\epsilon^{hom.}$ and $\epsilon_z=-2 \nu_r\epsilon^{hom.}/(1-\nu_r)$; 
all other components of the strain field are zero. Clearly, imposing a radial stretching also causes {\it nonzero tangential and vertical strains}.
The resulting stresses are Eqs. (\ref{hom1}-\ref{hom3}).

The case of the angular-dependent component is much more complex.
Injecting in the above the {\it ansatz} Eqs. (\ref{equaur}-\ref{equauz}) for the displacement fields, and writing the problem in reduced coordinates, we realize that the solution should be of the type Eqs. (\ref{fr}-\ref{fz}) at lowest order in $h/R_d$. The symmetry of the drum with respect to $z \rightarrow -z$ has been used.
To further reduce the problem, another {\it ansatz} is needed for the $\tilde{r}$-dependent functions introduced in the writing of the solution:
we assume them to be power laws, Eqs. (\ref{eq1}-\ref{eqn}).
Taking into account the boundary conditions (no $z$-component stress on the surface of the drum, and fixed radial strain at the periphery), we end up with the constants listed in Tab. \ref{tab2}.

The $\epsilon_0$ term is simply the prefactor of the angular-dependent strain, $\epsilon^{angl.}=\epsilon_0 \cos(2n \, \theta)$.
The stresses do depend on $\tilde{z}^2$. However, in the limit $h/R_d \rightarrow 0$ these terms vanish and the stress components are homogeneous within the thickness of the drum.
Also $\sigma_z=0$: the stress state is {\it planar}. The two normal components $\sigma_r^{angl.}, \sigma_\theta^{angl.}$ Eqs. (\ref{angl1}-\ref{angl2}) are displayed in Fig. \ref{fig_stress} for mode $\left\{n=2,m=1\right\}$ in the high-stress limit.

Furthermore, the only nonzero shear stress component is $\sigma_{r,\theta}^{angl.}$. It then writes:
\begin{displaymath}
\sigma_{r,\theta}^{angl.} = -E_r \frac{n (-3+\nu_r) +(-2+\nu_r)}{(1+2n)(1+\nu_r)} \left( \frac{r}{R_d}\right)^{2 n} \epsilon_0 \sin (2n \, \theta),
\end{displaymath}
with the $-$ sign matching our stress convention (tensile).
It is neglected in the presented modeling.

\begin{table}[h!]
\begin{center}
\begin{tabular}{|c|c|}    \hline
Parameter                     &  Expression                            \\    \hline    \hline
$ b_{r,0} $         &  $\epsilon_0/(1+2n)$                             \\    \hline
$ b_{\theta,0} $    &  $b_{r,0} \, 2(1+n) (2-\nu_r) /(2 n)^2$          \\    \hline
$ c_{r,0}      $    &  $0$                                   \\    \hline 
$ c_{\theta,0} $    &  $0$                                   \\    \hline
$ b_{z,0}      $    &  $0$                                   \\    \hline
$ c_{z,0}      $    &  $b_{r,0} \, 2(1+n) \nu_r$                        \\    \hline
$ a_{r,0}      $    &  $-b_{r,0} \, 2 [(1+n)(2-\nu_r)-n^2]$             \\    \hline
$ a_{\theta,0} $    &  $b_{r,0} \, n [1-4(1+n)(2-\nu_r)/(2n)^2]$        \\    \hline
$ a_{z,0}      $    &  $0$                                  \\    \hline 
$\alpha$            &  $2n+1$                                            \\    \hline 
$ \epsilon_{0} $    &  $\left( \frac{z_{n,m}}{R_d} \right)^2  \frac{ C_{n,m}^{(1)}-C_{n,m}^{(2)} }{2}$    \\    \hline 
\end{tabular}
\caption{\label{tab2} Strain coefficients of the angular-dependent contribution, as a function of $n, \nu_r$. $\epsilon_0$ is the amplitude of the $\cos(2n\, \theta)$ stretching term.}
\end{center}
\end{table}

\section{Mass, spring and Duffing parameters}
\label{modeparams}

In this Appendix we give numerical estimates for mass, spring constant and nonlinear parameters calculated for the first modes, in the two simple limits of high-stress and low-stress.

\begin{table}[h!]
\begin{center}
\begin{tabular}{|c|c|c|c|c|c|}    \hline
mode     $\left\{n,m\right\}$    &  H.S. $M_{n,m}$        &  H.S. $K_{n,m}$ & & L.S. $M_{n,m}$  &    L.S. $K_{n,m}$  \\    \hline    \hline
 $\left\{0,0\right\}$    &  0.269513       &  0.779325       & &  0.182834   &  9.54057                        \\    \hline
 $\left\{0,1\right\}$    &  0.115780       &  1.763983       & &  0.101896   &  80.5872                        \\    \hline
 $\left\{1,0\right\}$    &  0.239561       &  1.758616       & &  0.184581   &  41.7156                        \\    \hline
 $\left\{1,1\right\}$    &  0.133016       &  3.273413       & &  0.119933   &  221.883                        \\    \hline
 $\left\{0,2\right\}$    &  0.073686       &  2.759075       & &  0.067543   &  268.132                        \\    \hline
 $\left\{2,0\right\}$    &  0.243735       &  3.214208       & &  0.200046   &  121.669                        \\    \hline
 $\left\{1,2\right\}$    &  0.092082       &  4.765268       & &  0.085466   &  616.168                        \\    \hline
 $\left\{2,1\right\}$    &  0.155586       &  5.511635       & &  0.142446   &  509.546                        \\    \hline
 $\cdots$                &  $\cdots$       &  $\cdots$       & &  $\cdots$   &  $\cdots$                        \\    \hline
\end{tabular}
\caption{\label{tab3} Mass and spring constant for the first modes (norm. integrals, see text). Left: high-stress (H.S.) and right: low-stress (L.S.).}
\end{center}
\end{table}

For this purpose, we re-write the relevant integrals in an adimensional form such that: 
\begin{displaymath}
{\cal M}_{n,m}  =  \rho h \pi R_d^2 \, M_{n,m} , 
\end{displaymath}
and:
\begin{eqnarray}
& & {\cal K}_{m,n}  =    \nonumber \\
& & \frac{ 2\pi R_d \left|T_{r,0}\right|}{R_d} \, K_{n,m} \nonumber \\
& & \mbox{or} \nonumber \\
& & \frac{D_r \, 2\pi R_d}{ R_d^3} \, K_{n,m} , \nonumber 
\end{eqnarray}
in the high-stress and low-stress limits, respectively. 
$\rho h \pi R_d^2$ is the mass of the drum (in kg), and $2\pi R_d \left|T_{r,0}\right|$ the force tensioning the device at the periphery (in N, equivalent to $S_z$ for the beam case, see Figs. \ref{fig_scheme} and \ref{fig_drum}). 
Similarly, the flexural rigidity times perimeter $D_r \, 2 \pi R_d$ replaces the product $E_z I_z$ of the Euler-Bernoulli modeling. 
Numerical values for $M_{n,m}$ and $K_{n,m}$ are listed in Tab. \ref{tab3}.
Note that the mass parameters $M_{n,m}$ obtained in both high-stress and low-stress limits are very close. 
Resonance frequencies are then given by Eq. (\ref{freqref}).

\begin{table}[h!]
\begin{center}
\begin{tabular}{|c|c|c|c|c|c|}    \hline
mode     $\left\{n,m\right\}$    &  H.S. $C_{n,m}^{(1)}$        &  H.S. $C_{n,m}^{(2)}$ & & L.S. $C_{n,m}^{(1)}$  &    L.S. $C_{n,m}^{(2)}$  \\    \hline    \hline
 $\left\{0,0\right\}$    &  0.389664       &  0.389664       & &  0.316669   &  0.316669                       \\    \hline
 $\left\{0,1\right\}$    &  0.881992       &  0.881992       & &  0.851698   &  0.851698                       \\    \hline
 $\left\{1,0\right\}$    &  1.139994       &  0.618625       & &  0.920002   &  0.630386                       \\    \hline
 $\left\{1,1\right\}$    &  2.60152        &  0.671898       & &  2.50136    &  0.682519                       \\    \hline
 $\left\{0,2\right\}$    &  1.37954        &  1.37954        & &  1.34492    &  1.34492                        \\    \hline
 $\left\{2,0\right\}$    &  1.62609        &  1.58811        & &  1.31374    &  1.62645                        \\    \hline
 $\left\{1,2\right\}$    &  4.07293        &  0.692365       & &  3.96830    &  0.696559                        \\    \hline
 $\left\{2,1\right\}$    &  3.71904        &  1.79259        & &  3.55652    &  1.83054                        \\    \hline
 $\cdots$                &  $\cdots$      &  $\cdots$        & &  $\cdots$   &  $\cdots$                  \\    \hline
\end{tabular}
\caption{\label{tab4} Nonlinear coefficients $C_{n,m}^{(1,2)}$ computed for the first modes. Left: high-stress (H.S.) and right: low-stress (L.S.). Note the specificity of $n=0$ modes (by definition $C_{0,m}^{(1)}=C_{0,m}^{(2)}$, no angular dependence of strain/stress). }
\end{center}
\end{table}

In Tab. \ref{tab4} we give the stretching $C_{n,m}^{(1,2)}$ constants (no units) calculated for the first modes. High-stress and low-stress cases are again presented; the obtained numerical values in the two limits are very similar. 
As an illustrative example, the stretching function $\epsilon$ calculated for mode $\left\{n=2,m=1\right\}$ in the high-stress limit is presented in Fig. \ref{fig_stretch} (in normalized units).

\begin{table}
\begin{center}
\begin{tabular}{|c|c|c|c|}    \hline
mode     $\left\{n,m\right\}$                      &  H.S. $\tilde{K}_{n,m}^{(1)}$       & H.S. $\tilde{K}_{n,m}^{(2)}$ &    H.S. $\tilde{K}_{n,m}^{(3)}$    \\    \hline    \hline
 $\left\{0,0\right\}$                             &  0.779325   & X    &  X                             \\    \hline
 $\left\{0,1\right\}$                             &  1.76398    & X    &  X                             \\    \hline
 $\left\{1,0\right\}$                             &  1.75862    & 0.352992  &  -0.0598902               \\    \hline
 $\left\{1,1\right\}$                             &  3.27341    & 0.578822  &  -0.0332539               \\    \hline
 $\left\{0,2\right\}$                             &  2.75908    & X    &  X                             \\    \hline
 $\left\{2,0\right\}$                             &  3.21421    & 0.421149  &  -0.0997277               \\    \hline
 $\left\{1,2\right\}$                             &  4.76527    & 0.817227  &  -0.0230206                \\    \hline
 $\left\{2,1\right\}$                             &  5.51164    & 0.607419  &  -0.0562541                \\    \hline
 $\cdots$                                         &  $\cdots$   & $\cdots$  &  $\cdots$                  \\    \hline
\end{tabular}
\caption{\label{tab5} Nonlinear Duffing coefficients $\tilde{K}_{n,m}^{(1,2,3)}$ (no units) computed for the first modes, high-stress (H.S.) limit. Note that $\tilde{K}_{n,m}^{(1)}=K_{n,m}$ (H.S.), Tab. \ref{tab3}. For $n=0$ modes, $\tilde{K}_{0,m}^{(2,3)}$ are irrelevant (X above).}
\end{center}
\end{table}

We finally propose numerical estimates for the nonlinear coefficients written as:
\begin{eqnarray}
\tilde{{\cal K}}_{n,m} & = & +\frac{ E_r \, h 2 \pi R_d}{R_d^3}   \times \nonumber \\
& &  \!\!\!\!\!\!\!\!\!\!\!\!\!\!\!\!\!\!\!\!\!\! \left[ \frac{ C_{n,m}^{(1)}+C_{n,m}^{(2)} }{2}  \, \eta^{(0)}(\nu_r) \tilde{K}_{n,m}^{(1)} \, + \right.   \nonumber \\
&\!\!\!\!\!\!\!\!\!\!  &\!\!\!\!\!\!\!\!\!\!\!\!\!\!\!\!\!\!\!\! \left. \frac{ C_{n,m}^{(1)}-C_{n,m}^{(2)} }{2}  \times  \left( \eta_r^{(n)}(\nu_r) \tilde{K}_{n,m}^{(2)}  + \eta_\theta^{(n)}(\nu_r) \tilde{K}_{n,m}^{(3)} \right) \right], \nonumber 
\end{eqnarray}
where the adimensional $\tilde{K}_{n,m}^{(1,2,3)}$ are given in Tabs. \ref{tab5} and \ref{tab5bis} (high-stress and low-stress limits respectively).
Note the chosen normalization, that matches the Euler-Bernoulli formalism with $ h 2 \pi R_d$ the cross-section area of the device at the clamp; in the high-stress limit, $\tilde{K}_{n,m}^{(1)}=K_{n,m}$.

From Tabs. \ref{tab5} and \ref{tab5bis}, the $C_{n,m}^{(1,2)}$ values of Tab. \ref{tab4} and the expressions of the functions $\eta_{r,\theta}^{(n)}(\nu_r)$ [Eqs. (\ref{etas1},\ref{etas2}) and subsequent text], one realizes that the geometrical Duffing nonlinear parameter is {\it dominated by the homogeneous contribution}.
As a result, $\tilde{K}_{n,m}$ is {\it always positive}, as in the beam case. 
Finally, one can see that the numerical evaluations of $\tilde{K}_{n,m}^{(1,2,3)}$ are about twice larger in the high-stress limit than in the low-stress case. 
As such, for identical material parameters ($E_r, \nu_r, \rho$) except the biaxial stress $\sigma_0$ and identical geometry ($R_d, h$), a membrane Duffing nonlinearity $\tilde{K}_{n,m}$ (H.S.) is approximately {\it twice larger} than for a plate (L.S.).

\begin{table}[h!]
\begin{center}
\begin{tabular}{|c|c|c|c|}    \hline
mode     $\left\{n,m\right\}$                      &  L.S. $\tilde{K}_{n,m}^{(1)}$       & L.S. $\tilde{K}_{n,m}^{(2)}$ &    L.S. $\tilde{K}_{n,m}^{(3)}$    \\    \hline    \hline
 $\left\{0,0\right\}$                             &  0.316669    &  X   &  X                             \\    \hline
 $\left\{0,1\right\}$                             &  0.851698    &  X   &  X                             \\    \hline
 $\left\{1,0\right\}$                             &  0.775194    &  0.205519  & -0.0461452                \\    \hline
 $\left\{1,1\right\}$                             &  1.59194     &  0.433603  & -0.0299831                \\    \hline
 $\left\{0,2\right\}$                             &  1.34492     &  X   &  X                              \\    \hline
 $\left\{2,0\right\}$                             &  1.47010     &  0.209512  & -0.066682                 \\    \hline
 $\left\{1,2\right\}$                             &  2.33243     &  0.663581  & -0.0213665                \\    \hline
 $\left\{2,1\right\}$                             &  2.69353     &  0.388931  & -0.0474821                \\    \hline
 $\cdots$                                         &  $\cdots$    & $\cdots$  & $\cdots$                   \\    \hline
\end{tabular}
\caption{\label{tab5bis} Nonlinear Duffing coefficients $\tilde{K}_{n,m}^{(1,2,3)}$ (no units) computed for the first modes, low stress (L.S.) limit. For $n=0$ modes, $\tilde{K}_{0,m}^{(2,3)}$ are irrelevant (X above).}
\end{center}
\end{table}

\end{document}